\documentclass[sigconf,natbib=true]{acmart}

\usepackage[normalem]{ulem}
 \usepackage{amsmath}
  \usepackage{multirow}
\usepackage{epstopdf}
  \usepackage{graphicx}
\usepackage{appendix}

\usepackage[normalem]{ulem}
\useunder{\uline}{\ul}{}
 \usepackage{enumitem}
 \useunder{\uline}{\ul}{}
\AtBeginDocument{%
  \providecommand\BibTeX{{%
    \normalfont B\kern-0.5em{\scshape i\kern-0.25em b}\kern-0.8em\TeX}}}

\copyrightyear{2024}
\acmYear{2024}
\setcopyright{rightsretained}
\acmConference[WSDM'24]{Proceedings of the 17th ACM International Conference on Web Search and Data Mining}{March 4th-8th, 2024}{Mérida, Yucatán, México}
\acmBooktitle{Proceedings of the 17th ACM International Conference on Web Search and Data Mining (WSDM'24), Mérida, Yucatán, México}
\acmDOI{10.1145/3543507.3583278}
\acmISBN{978-1-4503-9416-1/23/04}

\usepackage{etoolbox}
\makeatletter
\patchcmd{\maketitle}{\@copyrightpermission}{
   \begin{minipage}{0.3\columnwidth}
     \href{http://creativecommons.org/licenses/by/4.0/}{\includegraphics[width=0.90\textwidth]{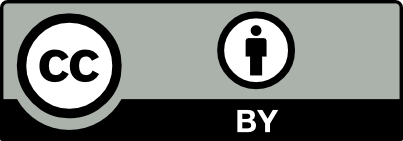}}
   \end{minipage}\hfill
   \begin{minipage}{0.7\columnwidth}
     \href{http://creativecommons.org/licenses/by/4.0/}{This work is licensed under a Creative Commons Attribution International 4.0 License.}
   \end{minipage}
  
   \vspace{5pt}
}{}{}

\author{Guanyu Lin$^{1}$, Chen Gao$^{2}$, Yu Zheng$^{2}$, Jianxin Chang$^{3}$, Yanan Niu$^{3}$, Yang Song$^{3}$, Kun Gai$^{5}$, Zhiheng Li$^{2}$, Depeng Jin$^{2}$, Yong Li$^{2}$, Meng Wang$^{4}$}
\affiliation{
 \institution{$^1$Carnegie Mellon University, $^2$Tsinghua University, $^3$Kuaishou Technology, $^4$Hefei University of Technology, $^5$Unaffiliated}
 \institution{guanyul@andrew.cmu.edu, chgao96@gmail.com, zhengyu.davy@foxmail.com, \\ \{changjianxin, niuyanan, yangsong\}@kuaishou.com,  gai.kun@qq.com, \\ \{zhhli, jindp, liyong07\}@tsinghua.edu.cn, eric.mengwang@gmail.com}
 \country{}
}

\renewcommand{\shortauthors}{Lin et al.}

\begin{document}

\title{Mixed Attention Network for Cross-domain Sequential Recommendation}

\begin{abstract}
In modern recommender systems, sequential recommendation leverages chronological user behaviors to make effective next-item suggestions, which suffers from data sparsity issues, especially for new users. One promising line of work is the cross-domain recommendation, which trains models with data across multiple domains to improve the performance in data-scarce domains. Recent proposed cross-domain sequential recommendation models such as PiNet and DASL have a common drawback relying heavily on overlapped users in different domains, which limits their usage in practical recommender systems. In this paper, we propose a \textbf{M}ixed \textbf{A}ttention \textbf{N}etwork (MAN) with local and global attention modules to extract the domain-specific and cross-domain information. Firstly, we propose a local/global encoding layer to capture the domain-specific/cross-domain sequential pattern. Then we propose a mixed attention layer with item similarity attention, sequence-fusion attention, and group-prototype attention to capture the local/global item similarity, fuse the local/global item sequence, and extract the user groups across different domains, respectively. Finally, we propose a local/global prediction layer to further evolve and combine the domain-specific and cross-domain interests. Experimental results on two real-world datasets (each with two domains) demonstrate the superiority of our proposed model. Further study also illustrates that our proposed method and components are model-agnostic and effective, respectively. The code and data are available at \url{https://github.com/Guanyu-Lin/MAN}.
\end{abstract}


\maketitle

\section{Introduction}\label{sec:intro}
Widespread in online platforms such as news, video, e-commerce, etc., recommender systems that vastly improve the efficiency of information distribution and diffusion are of great importance in today's Web.
Sequential recommendation~\cite{SRs} is one of the most important research problems in recommender systems, which aims at predicting a user's next interacted item based on their historical interaction sequence.
Though recent representative models of sequential recommendation such as GRU4REC~\cite{GRU4REC}, SASRec~\cite{kang2018self} and SURGE~\cite{SURGE} etc. have achieved decent performance,  
they suffer from the issue of data sparsity ~\cite{wang2021counterfactual},
limiting the performance.

\begin{figure}[t!]
	\begin{center}
			\includegraphics[width=.97\linewidth]{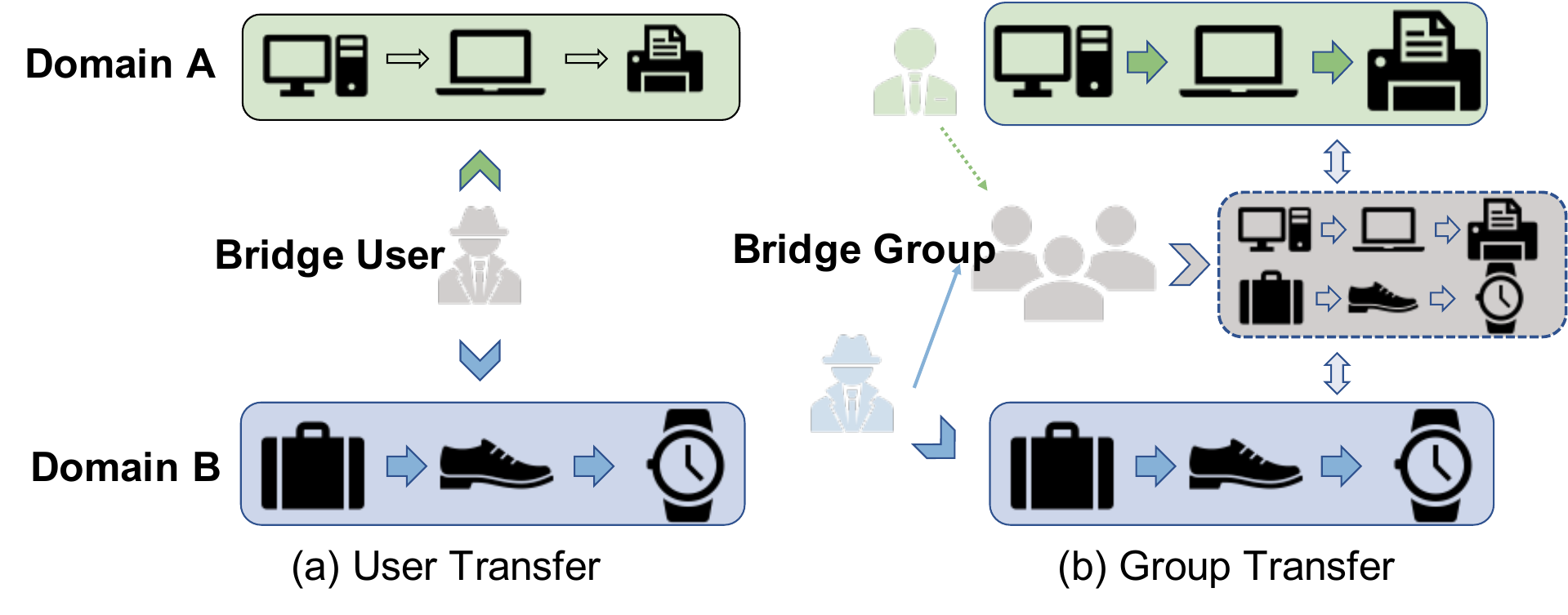} 
	\end{center}
	\vspace{-7px}
	\caption{Illustration of (a) user transfer learning relies on overlapped users and (b) group transfer learning without previous assumptions on user overlap.}	\label{fig:overlap}
	\vspace{-0.5cm}
\end{figure} 
To address the data sparsity issue, cross-domain recommendation~\cite{CATN, Bridge, itemCST} is a widely adopted approach, which leverages the data from multiple domains to boost the performance of the data-scarce domain by parameter-sharing~\cite{Bridge} or multi-task learning~\cite{lu2018like}.
Particularly, a few early attempts~\cite{PiNet, DASL,chen2021dual} were proposed to achieve {\em cross-domain sequential recommendation}, which leverages cross-domain technique to address the data sparsity of sequential modeling.
However, as illustrated in Figure~\ref{fig:overlap}(a), these methods rely heavily on the overlapped users and require pairwise inputs from two domains of the same bridge users, which is hardly satisfied in practical scenarios. 
For example, in our experimental benchmark datasets (Micro Video and Amazon), there is only a small part (at most 8.37\%) of overlapped users, as Table~\ref{tbl:data_mv}, which violates the assumption of existing approaches.
In fact, in many real-world applications, users are not overlapping across different domains~\cite{lin2020fedrec, liang2021fedrec++}. Thus, it is challenging for the existing cross-domain sequential recommendation to work in real-world scenarios. Actually, there are three key challenges for cross-domain sequential recommendation:
\begin{itemize}[leftmargin=*]
    \item \textbf{Different item characteristics across domains.}
    There are not always overlapped items across different domains. Even if items are shared across domains, items reflect different characteristics. For example, for a higher-end e-commerce website, the price aspect takes less effect when users purchase items, while it plays an important role in a lower-end website. Such difference brings difficulty in learning accurate item representations across different domains.
    \item \textbf{Various sequential patterns across domains.} 
    Similar to the item, the sequential behaviors vary in different domains. For example, users may be more decisive in a higher-end E-commerce website, leading to very short sequences with very brief sequential patterns. Therefore sequential patterns are various, and the modeling is challenging.
    \item \textbf{User preference transferring without overlapped user.} We focus on the general cross-domain recommendation task, where users may not fully overlap.
    Therefore, it is challenging to capture the common preference shared by users across domains, especially when there is even no overlapped user.
\end{itemize}
To address these challenges, we develop a novel group-based method with the group transfer to avoid dependence on the overlap of users and global space to capture the item characteristics and sequential patterns across different domains as Figure~\ref{fig:overlap}(b). Note that the group-prototype attention here can capture group information in an unsupervised manner, without further requiring additional input information compared with Figure~\ref{fig:overlap}(a).
Specifically, we propose a novel solution named MAN (short for \textbf{M}ixed \textbf{A}ttention \textbf{N}etwork for Cross-domain Sequential Recommendation), consisting of local and global modules, mixing three types of designed attention network from item level, sequence level, and group level. 
First, we generate separate representations for each item, including the local representation capturing domain-specific characteristics and the global representation shared by different domains. We then design an item similarity attention module to capture the similarity between local/global item representation and the target item representation.
Second, we propose a sequence-fusion attention module to fuse the local and global item sequential representations.
Most importantly, although user information cannot be directly shared, the group information can be shared across domains.
Therefore, we propose a group-prototype attention module, which utilizes multiple group prototypes to transfer the information at the group level. 
Finally, the obtained local and global embeddings are fed into the corresponding prediction layers to evolve the domain-specific and cross-domain interests. 

The contributions of this paper can be summarized as follows.

\begin{itemize}[leftmargin=*] 
    \item We approach the problem of cross-domain sequential recommendation from a more practical perspective that there is no prior assumption of overlapped users across domains, which is far more challenging.
    \item We propose a solution named MAN and address the key challenges by mixing three attention modules: item similarity attention, sequence-fusion attention, and group-prototype attention. Besides, local and global designs are proposed to capture the domain-specific and cross-domain patterns.
    \item We conduct extensive experiments on a collected large-scale industrial dataset and a public benchmark dataset, where the results show significant performance improvements compared with the state-of-the-art models. Further studies illustrate that our proposed method is model-agnostic, and group prototypes can capture the group patterns across domains without overlapping users.
\end{itemize}

\section{Problem Formulation}


    

In our problem of cross-domain sequential recommendation, we first use A and B to denote the two domains.
Let $\mathcal{I}^{A}$ and $\mathcal{I}^{B}$ denote the sets of items in domain A and B, respectively.
More specifically, supposing $i^{A}_{t} \in \mathcal{I}^{A}$ or $i^{B}_{t} \in \mathcal{I}^{B}$ is the $t$-th item that a given user has interacted with in the A or B domain, the $t$-length sequence of historical items can be represented as $(i^A_{1}, i^A_{2}, \ldots, i^A_{t})$ or $(i^B_{1}, i^B_{2}, \ldots, i^B_{t})$. 
The goal of our problem is to improve the recommendation accuracy of the following item \textit{i.e.}, $i^A_{t+1}$ or $i^B_{t+1}$, of all users across all domains simultaneously. 
The problem can be formulated as follows. 

    \noindent \textbf{Input}: Item sequence $(i^A_{1}, i^A_{2}, \ldots, i^A_{t} )$ and $(i^B_{1}, i^B_{2}, \ldots, i^B_{t} )$ for users in domain A and B, respectively.
    
    \noindent \textbf{Output}: The cross-domain recommendation model estimating the probability that target item $i^A_{t+1}$ and $i^B_{t+1}$ will be interacted by the given users with item sequence $ (i^A_{1}, i^A_{2}, \ldots, i^A_{t})$ and $ (i^B_{1}, i^B_{2}, \ldots, i^B_{t})$ in the domain A and B, respectively.

\section{Methodology}
\begin{figure*}[t!]
	\begin{center}
		\begin{tabular}{c}
		\includegraphics[width=.97\linewidth]{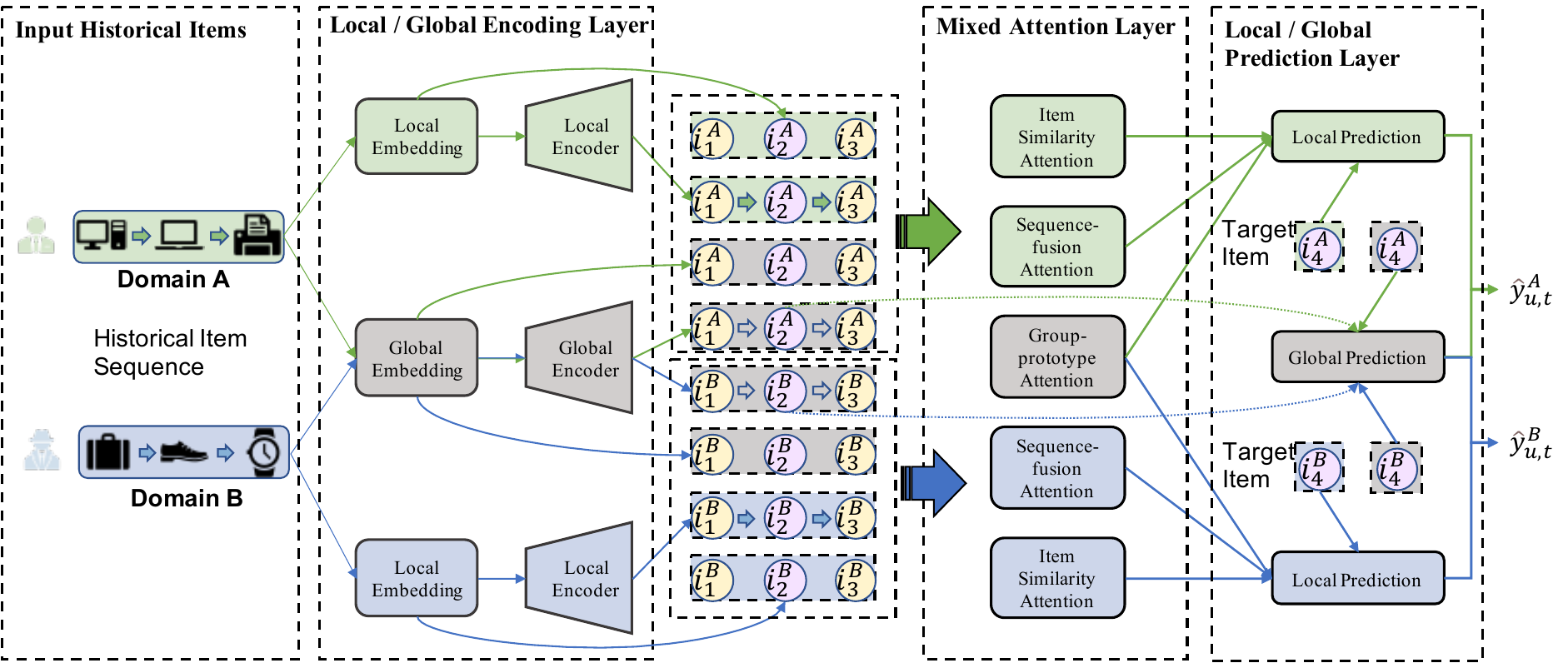} \\
		\includegraphics[width=.97\linewidth]{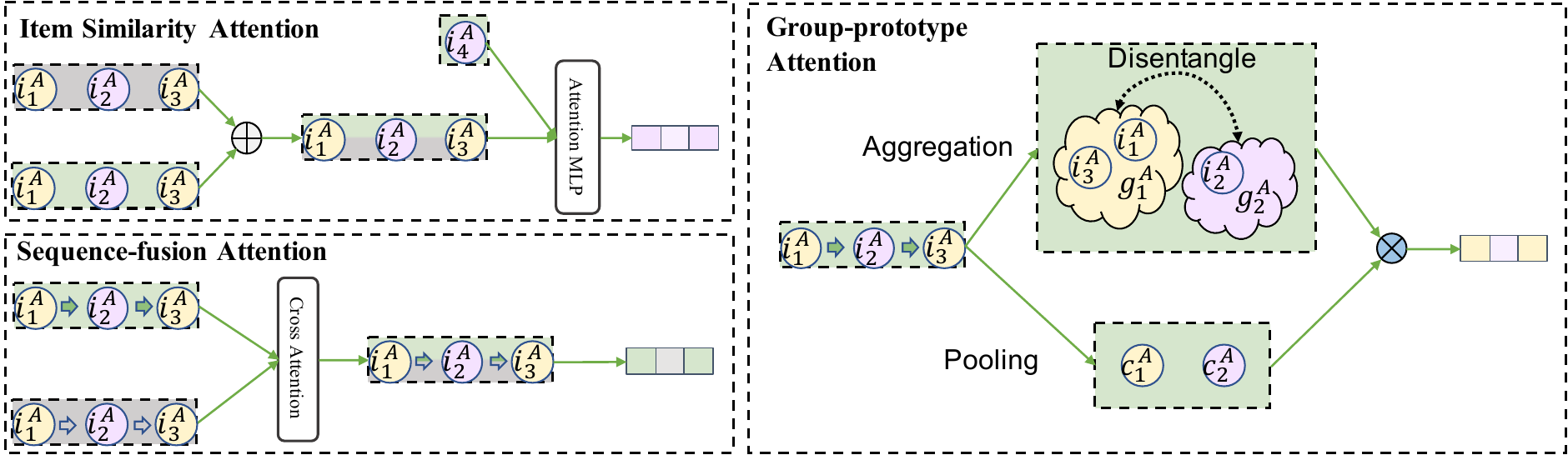}
		\end{tabular}
	\end{center}
	\caption{Illustration of our proposed MAN model. (1) The item sequences are first input into the Local/Global Encoding Layer, which builds local and global embeddings for each item and encodes them to extract the local and global sequential patterns; (2) In the Mixed Attention Layer, Item Similarity Attention is fed with local and global item embeddings to capture the item-level relation; Sequence-fusion Attention fuses the encoded local and global sequential representations to capture the sequence-level relation; Group-prototypes attention leverages the shared group prototypes to capture the group-level relation. Here we take domain A to illustrate each proposed attention component in detail.
	(4) The aggregated embeddings will be fed into the local prediction layer and global prediction layer, respectively, for the final prediction.}	\label{fig:architecture}
\end{figure*} 

Figure~\ref{fig:architecture} illustrates our proposed MAN model, encoding the item sequence with local/global encoding layer, mixing three attention modules, and evolving the interests by local/global prediction layer.

\begin{itemize}[leftmargin=*]

\item \textbf{Local/Global Encoding Layer}.
We build both domain-specific local and cross-domain global embeddings for items. We further encode them with local and global encoders, respectively, to capture the domain-specific and cross-domain item sequential patterns.


\item \textbf{Mixed Attention Layer}.
We propose Item Similarity Attention, Sequence-fusion Attention, and Group-prototype Attention to capture the cross-domain patterns at the item, sequence, and group levels.



 \item \textbf{Local/Global Prediction layer}. 
To evolve the interests and predict the probability of the candidate's next item that the user will interact with in each domain, we propose a local prediction layer and a global prediction layer.

\end{itemize}

\subsection{Local/Global Encoding Layer}
We first build local and global item embeddings. Then we further look up item embeddings and encode them with local and global encoders at the sequence level. 

\subsubsection{\textbf{Local and Global Item Embeddings}}

To capture the domain-specific patterns for different domains, we create two item embedding matrices $\textbf{M}^A \in \mathbb{R}^{|\mathcal{I}^A | \times D}$ and $\textbf{M}^B \in \mathbb{R}^{|\mathcal{I}^B | \times D}$ where $D$ denotes the latent dimensionality.
Then, to capture the shared item characteristics across different domains, from the perspective of representation learning, we assume there exists a shared latent space~\cite{wang2019heterogeneous} where different domains have common representation; Thus, we create a shared embedding matrix $\textbf{M}\in \mathbb{R}^{|\mathcal{I}^A \cup \mathcal{I}^B | \times D'}$.

Here we use $(i^A_{1}, i^A_{2}, \ldots, i^A_{t})$ and $(i^B_{1}, i^B_{2}, \ldots, i^B_{t})$, to denote the historical item sequences of domain A and domain B, specifically.
Note that we pad sequences shorter than $T$ with a constant zero vector, following existing works~\cite{kang2018self, SURGE}. 
To further capture the position of items in the sequence, we also integrate learnable \textit{positional embeddings} into item embeddings (domain A example) as:

\begin{equation}\label{eqn::emb}
\small
\mathbf{E}^{A}=\left[\begin{array}{c}
	\textbf{M}^A_{i^A_{1}}+\textbf{P}^A_{1} \\
	\textbf{M}^A_{i^A_{2}}+\textbf{P}^A_{2} \\
	\cdots \\
	\textbf{M}^A_{i^A_{n}}+\textbf{P}^A_{n}
\end{array}\right],
\mathbf{E}^{A_g}=\left[\begin{array}{c}
	\textbf{M}_{i^{A}_{ 1}}+\textbf{P}_{1} \\
	\textbf{M}_{i^{A}_{ 2}}+\textbf{P}_{2} \\
	\cdots \\
	\textbf{M}_{i^{A}_{ n}}+\textbf{P}_{n}
\end{array}\right],
\end{equation}
\begin{equation}
\small
\mathbf{E}^{B}=\left[\begin{array}{c}
	\textbf{M}^{B}_{i^{B}_{1}}+\textbf{P}^{B}_{1} \\
	\textbf{M}^{B}_{i^{B}_{2}}+\textbf{P}^{B}_{2} \\
	\cdots \\
	\textbf{M}^{B}_{i^{B}_{n}}+\textbf{P}^{B}_{n}
\end{array}\right],
\mathbf{E}^{B_g}=\left[\begin{array}{c}
	\textbf{M}_{i^{B}_{ 1}}+\textbf{P}_{1} \\
	\textbf{M}_{i^{B}_{ 2}}+\textbf{P}_{2} \\
	\cdots \\
	\textbf{M}_{i^{B}_{ n}}+\textbf{P}_{n}
\end{array}\right],
\end{equation}
where $\mathbf{E}^{A}$, $\mathbf{E}^{B} \in \mathbb{R}^{T \times D}$ ($\mathbf{E}^{A_g}$, $\mathbf{E}^{B_g} \in \mathbb{R}^{T \times D'}$) denote the local (global) embeddings for domain $A$ and domain $B$, respectively. Besides, $g$ means global.
Here $\textbf{P}^A$, $\textbf{P}^B \in \mathbb{R}^{T \times D}$ and $\textbf{P} \in \mathbb{R}^{T \times D'}$ are the learnable positional embeddings. 

\subsubsection{\textbf{Local Encoder and Global Encoder of Sequences}}

After obtaining $\mathbf{E}^{A}$, $\mathbf{E}^{B}$, $\mathbf{E}^{A_g}$ and $\mathbf{E}^{B_g}$ from the embedding layers, we then apply sequential encoders to learn the sequential patterns.
Here we propose the local encoder and global encoder as follows,
\begin{equation}\label{eq:localA} 
	\mathbf{S}^A= \textbf{Encoder}(\mathbf{E}^A),
	\mathbf{S}^B= \textbf{Encoder}(\mathbf{E}^{B}),
\end{equation}
	\begin{equation}\label{eq:globalA} 
	\mathbf{S}^{A_g}= \textbf{Encoder}_g(\mathbf{E}^{A_g}),
	\mathbf{S}^{B_g}= \textbf{Encoder}_g(\mathbf{E}^{B_g}),
\end{equation}
where $\textbf{Encoder}$ and $\textbf{Encoder}_g$ are the sequential backbone models (i.e., SASRec~\cite{kang2018self} or SURGE~\cite{SURGE}) with independent and shared parameters, respectively, across domains\footnote{Note that all functions are with dependent parameters except that they are subscript with $g$ w.r.t. $global$.}.

Based on it, we obtain $\mathbf{S}^A$ ($\mathbf{S}^B$) and $\mathbf{S}^{A_g}$ ($\mathbf{S}^{B_g}$), which capture local sequential patterns and global sequential patterns, respectively, in domain A (B).
\subsection{Mixed Attention Layer}
In this section, we first propose item similarity attention to extract similar items from local and global spaces. Then we propose sequence-fusion attention to further fuse the local and global item sequence representations, which will combine the domain-specific and cross-domain sequential patterns. Finally, we propose group-prototype attention to extract the group pattern across domains.
\subsubsection{\textbf{Item Similarity Attention}}

To capture the similarity between local/global item embeddings and target item embedding, we first fuse the item embedding from local space (i.e., $\mathbf{E}^{A}_{j}$ and $\mathbf{E}^{B}_{j}$) and global space (i.e., $\mathbf{E}^{A_g}_{j}$ and $\mathbf{E}^{B_g}_{j}$ ) together.
Specifically, given a user in domain A (B), we can calculate the item similarity scores $\mathbf{F}^A$ ($\mathbf{F}^B$) between his/her historical items and the target item as follows,
\begin{equation}\label{eq:similarScore_A}
	\mathbf{F}^A=\text{MLP}\left( \textbf{M}_{i^A_{t+1}} \| \mathbf{E}^{A} + \mathbf{E}^{A_g} \right) ,
	\mathbf{F}^B=\text{MLP}\left( \textbf{M}_{i^B_{t+1}} \| \mathbf{E}^{B} + \mathbf{E}^{B_g}\right) ,
\end{equation}
 where $\mathbf{M}_{i^A_{t+1}}$ ($\mathbf{M}_{i^B_{t+1}}$) denotes the embedding of the target item for domain A (B) and $\|$ denotes the concatenation operation. Based on the item similarity scores, we can then weigh similar historical items' embeddings to refine item embeddings as follows,
\begin{equation}\label{eq:similarItem} \small
	\mathbf{E}^{A_i}=\textbf{softmax}\left( \mathbf{F}^{A}\right)(\mathbf{E}^{A} + \mathbf{E}^{A_g} ) ,
	\mathbf{E}^{B_i}=\textbf{softmax}\left( \mathbf{F}^{B}\right)(\mathbf{E}^{B} + \mathbf{E}^{B_g}) ,
\end{equation}
where $\mathbf{E}^{A_i}$ and $\mathbf{E}^{B_i} \in \mathbb{R}^{T \times D}$ are the representations of target items' similar historical items weighted by similarity scores of $\mathbf{F}^{A}$ and $\mathbf{F}^{B}$ in domain A and domain B, respectively. Here ${A_i}$ and ${B_i}$ mean item similarity of domain $A$ and $B$, respectively.

\subsubsection{\textbf{Sequence-fusion Attention}}\label{sec:chronological}

After obtaining $\mathbf{S}^A$, $\mathbf{S}^B$, $\mathbf{S}^{A_g}$, and $\mathbf{S}^{B_g}$,
we then fuse them to combine the domain-specific and cross-domain sequential patterns together as follows,
\begin{equation} \small
	\mathbf{S}^{A_s}= \textbf{MLP} (\textbf{CA} ( \mathbf{S}^A, \mathbf{S}^{A_g}) + \mathbf{S}^A);
\mathbf{S}^{B_s}= \textbf{MLP} (\textbf{CA} ( \mathbf{S}^B, \mathbf{S}^{B_g}) + \mathbf{S}^B);
\end{equation}
where the \textbf{c}ross-\textbf{a}ttention (\textbf{CA}) layer~\cite{vaswani2017attention} is defined as follows (take $\mathbf{S}^A$ as an example),
\begin{equation}\label{eq:ca}
    \textbf{CA}(\mathbf{S}^A, \mathbf{S}^{A_g})  = \textbf{Atten}\left(\mathbf{S}^A \textbf{W}^{Q}_{A_s}, \mathbf{S}^{A_g} \textbf{W}^{K}_{A_s}, \mathbf{S}^{A_g} \textbf{W}^{V}_{A_s}\right),
\end{equation}
where $\mathbf{W}^{Q}_{A_s}$, $\mathbf{W}^{K}_{A_s}$, $\mathbf{W}^{V}_{A_s} \in \mathbb{R}^{D \times D}$ are parameters to be learned and $\textbf{Atten}$ function is defined as below.
\begin{equation}\label{eq:dotAtten}
	\textbf{Atten}(\textbf{Q}, \textbf{K}, \textbf{V})=\textbf{softmax}\left(\frac{\textbf{Q K}^{T}}{\sqrt{D}}\right) \textbf{V}, 
\end{equation}
where $\textbf{Q}$, $\textbf{K}$, and $\textbf{V}$ are the query matrix, key matrix, and value matrix, respectively.

\subsubsection{\textbf{Group-prototype Attention}}\label{sec:tsa}

Although we can not leverage overlapped user IDs across domains, there often exist user groups with similar preferences. Specifically, we first pool each sequence to obtain relevance to each group. Then we leverage multiple group prototypes to aggregate the item groups and weigh them based on their relevance.


\paragraph{\textbf{Group Interest Pooling}}
For an item sequence of a user, it actually does not belong to only one group prototype. Instead, it can be a hybrid combination of several prototypes with different weights. For example, a user can be both an adolescent and a basketball lover at the same time.
Thus, we propose a learnable soft cluster assignment matrix~\cite{ying2018hierarchical, ranjan2020asap}, to calculate the importance of $N_g$ groups.
Specifically, the item sequence of each user is firstly pooled by a pooling matrix $\textbf{W}^{P}_A \in \mathbb{R}^{N_g \times T}$ ( $\textbf{W}^{P}_B \in \mathbb{R}^{N_g \times T}$), based on which the relevance of the user to each group can be calculated as follows, 
\begin{equation}\label{eq:cluster_a}
\mathbf{C}^{A} =  \textbf{MLP}\left(\textbf{W}^{P}_A \mathbf{S}^A\right),
	\mathbf{C}^{B} = \textbf{MLP}\left(\textbf{W}^{P}_B \mathbf{S}^B \right),
\end{equation}
where $\mathbf{C}^{A}$ and $\mathbf{C}^{B} \in \mathbb{R}^{N_g \times 1}$ are relevance scores for each group. 

\paragraph{\textbf{Group Interest Aggregation}}
We then create $N_g$ group-prototype embeddings $\mathbf{G} \in \mathbb{R}^{N_g \times D}$ to represent the interest groups. 
These embeddings can then be transformed to each domain, aggregating the typically related items as follows,
\begin{equation}
		\mathbf{G}^{A}= \textbf{MLP}\left(\textbf{CA}(\mathbf{G}, \mathbf{S}^A)\right),
	\mathbf{G}^{B}= \textbf{MLP}\left(\textbf{CA}(\mathbf{G}, \mathbf{S}^B)\right),
\end{equation}
where $\mathbf{G}^{A}$ and  $\mathbf{G}^{B} \in \mathbb{R}^{N_g \times D}$ are the obtained group-prototype representations for the sequences of domain A and domain B, respectively. Here \textbf{CA} layer is similar to Eqn.\eqref{eq:ca}.
We can then weigh all group-prototype representations based on the relevance scores as follows,
\begin{equation}\label{eq:group_repre_a}
	\mathbf{G}^{A_u}= \textbf{softmax}\left(\mathbf{C}^{A}\right) \mathbf{G}^{A}, 
	\mathbf{G}^{B_u}= \textbf{softmax}\left(\mathbf{C}^{B}\right) \mathbf{G}^{B} ,
\end{equation}
where $\mathbf{G}^{A_u}$ and $\mathbf{G}^{B_u} \in \mathbb{R}^{N_g \times D}$ are the weighted group-prototype representations for each user.

\paragraph{\textbf{Group-prototype Disentanglement}}

Each group prototype obviously should be distinct, according to its definition.
Therefore, inspired by the advances of disentangled representation learning~\cite{ma2020disentangled}, we propose the prototype disentanglement regularization as: 
\begin{equation}\label{eq:disen}
\mathcal{L}^{g}= - \lambda_g \sum_{i=1}^{N_g} \sum_{j=i + 1}^{N_g} \left(\mathbf{G}_i - \mathbf{G}_j\right)^2
\end{equation}
where $\lambda_g$ is the penalty hyper-parameter.
This loss function will be jointly learned with the main loss function later.

\subsection{Local/Global Prediction Layer}
In this section, we first evolve the local and global interests via corresponding prediction layers. Then we optimize them with the objective function for each domain.
\subsubsection{\textbf{Local and Global Prediction Layer}}

With the proposed mixed-attention network (item similarity attention, sequence-fusion attention, and group-prototype attention), we concatenate the outputs together and feed them into the proposed local prediction layer and global prediction layer based on MLP~\cite{DIN, DIEN, SURGE}, which can be formulated as follows,
\begin{equation}
	\hat{y}^A_{u, t}=\textbf{MLP}\left(\mathbf{e}^{A_i} \| \mathbf{s}^{A_s}\|\mathbf{g}^{A_u} \| \mathbf{s}^{A} 	 \| \textbf{M}^A_{i^A_{t+1}}\right) + \textbf{MLP}_g\left(\mathbf{s}^{A_g} \| \textbf{M}_{i^A_{t+1}}\right),
\end{equation}
\begin{equation}
	\hat{y}^B_{u, t}=\textbf{MLP}\left(\mathbf{e}^{B_i} \| \mathbf{s}^{B_s}\|\mathbf{g}^{B_u} \| \mathbf{s}^{B} 	 \| \textbf{M}^B_{i^B_{t+1}}\right) + \textbf{MLP}_g\left(\mathbf{s}^{B_g} \| \textbf{M}_{i^B_{t+1}}\right),
\end{equation}
where $\textbf{MLP}_g$ is the MLP layer with shared parameters across domains, and the concatenated embeddings are obtained via,
$$ \small
\begin{aligned}
\mathbf{e}^{A_i}= \sum_{t=1}^{T} \mathbf{E}^{A_i}_t, \mathbf{s}^{A_s} = \sum_{t=1}^{T} \mathbf{S}^{A_s}_t, \mathbf{g}^{A_u}= \sum_{k=1}^{N_g} \mathbf{G}^{A_u}_k,  \mathbf{s}^{A} = \sum_{t=1}^{T}\mathbf{S}^A_t, \mathbf{s}^{A_g} = \sum_{t=1}^{T}\mathbf{S}^{A_g}_t, \\
\mathbf{e}^{B_i}= \sum_{t=1}^{T} \mathbf{E}^{B_i}_t,\mathbf{s}^{B_s} = \sum_{t=1}^{T}\mathbf{S}^{B_s}_t, 
\mathbf{g}^{B_u}= \sum_{k=1}^{N_g} \mathbf{G}^{B_u}_k, 
\mathbf{s}^{B} = \sum_{t=1}^{T}\mathbf{S}^B_t, \mathbf{s}^{B_g} = \sum_{t=1}^{T}\mathbf{S}^{B_g}_t,
 \end{aligned} 
$$
which denotes average pooling before being fed into MLPs. 

\subsubsection{\textbf{Objective Function with Independent Updating}}
We then exploit the negative log-likelihood function~\cite{DIN, DIEN, SURGE} for optimization, which can be formulated as follows, 
\begin{equation}\label{eq:lossA}\small
\mathcal{L}^A=-\frac{1}{|\mathcal{R}^A|} \sum_{(u, i^A_t) \in \mathcal{R}^A}\left(y^A_{u, t} \log \hat{y}^A_{u, t}+\left(1-y^A_{u, t}\right) \log \left(1-\hat{y}^A_{u, t}\right)\right),
\end{equation}

\begin{equation}\label{eq:lossB}\small
\mathcal{L}^B=-\frac{1}{|\mathcal{R}^B|} \sum_{(u, i^B_t) \in \mathcal{R}^B}\left(y^B_{u, t} \log \hat{y}^B_{u, t}+\left(1-y^B_{u, t}\right) \log \left(1-\hat{y}^B_{u, t}\right)\right),
\end{equation}
where $\mathcal{R}^A$ and $\mathcal{R}^B$ are the training sets of domain A and domain B, respectively. 
Here $y^A_{u, t}=1$ ($y^B_{u, t}=1$) and $y^A_{u, t}=0$ ($y^B_{u, t}=0$) indicate a positive sample and a negative sample, respectively, and $\hat{y}^A_{u, t}$ and $\hat{y}^B_{u, t}$ stand for predicted click probability of the next item.

To optimize jointly across two domains, the final objective function is a linear combination of $\mathcal{L}^A$, $\mathcal{L}^B$ and $\mathcal{L}^{g}$ calculated in Eqn.\eqref{eq:disen}, Eqn.\eqref{eq:lossA} and Eqn.\eqref{eq:lossB}, respectively, as follows,
\begin{equation}\label{eq:loss}
\mathcal{L}= \mathcal{L}^A + \mathcal{L}^B + \lambda^A\|\Theta^A\|_{2} + \lambda^B\|\Theta^B\|_{2} + \mathcal{L}^{g}
\end{equation}
where $\Theta^A$ and $\Theta^B$ are the sets of learnable parameters with $\lambda^A$ and $\lambda^B$ as the regularization penalty hyper-parameters of domain A and domain B, respectively.

\noindent \textit{\textbf{Discussion.}}
Different from the existing works of cross-domain sequential recommendation such as PiNet~\cite{PiNet} and DASL~\cite{DASL} that are based on bridge users, our proposed MAN model does not rigidly require item sequences from two domains as input at the same time, since our proposed model's output of each domain does not require the input of another domain. 
That is to say, each domain in our model can update its parameters independently. 
If there is no input from any domain, we can easily just remove the optimization goal of that domain, \textit{e.g.} the loss function of Eq.\eqref{eq:loss} will be simplified as $\mathcal{L}=  \mathcal{L}^B + \lambda^B\|\Theta^B\|_{2} + \mathcal{L}^{g}$ if there is no input from domain A. 
In the real world's online recommendation, our MAN is more practical since the newly collected data from two domains are always not synchronous (our MAN can be optimized iteratively for each domain).

\section{Experiments}
In this section, we conduct extensive experiments with two real-world datasets, investigating the following research questions (RQs).
\begin{itemize}[leftmargin=*]
	\item \textbf{RQ1:} How does the proposed method perform compared with the state-of-the-art single-domain recommenders and cross-domain recommenders?
		\item \textbf{RQ2:}  What is the effect of different components in the method? 
\item \textbf{RQ3:}  Is the proposed method model-agnostic? What about the performance on different backbones? Is the method still effective with the solely local or global module?
  \item \textbf{RQ4:}  How do the group prototypes represent different groups? 
\end{itemize}
We also study RQ5: "What is the optimal number of group prototypes?" in Appendix~\ref{appendix::group_num}.

\subsection{Experimental Setup}
\subsubsection{\textbf{Datasets}}


\begin{table}[!htb]
\caption{Data statistic for two datasets. Here Avg. Length is the average number of users' history interacted items.}\label{tbl:data_mv}
\small
\setlength\tabcolsep{2pt}   

\begin{tabular}{c|c|c|c|c}
\hline
\textbf{Dataset}       & \multicolumn{2}{c|}{\textbf{Micro Video}} & \multicolumn{2}{c}{\textbf{Amazon}}  \\ \hline
\textbf{Domain}        & \textbf{A}          & \textbf{B}         & \textbf{Video Games} & \textbf{Toys} \\ \hline
\textbf{\#Users}       & 43,919              & 37,692             & 826,767              & 1,342,911     \\ \hline
\textbf{\#Items}       & 147,813             & 131,732            & 50,210               & 327,698       \\ \hline
\textbf{\#Records}     & 18,011,737          & 14,908,625         & 1,324,753            & 2,252,771     \\ \hline
\textbf{Overlap Items} & 71.22\%             & 79.91\%            & 7.66\%               & 4.72\%        \\ \hline
\textbf{Overlap users} & 7.18\%              & 8.37\%             & 0.27\%               & 0.04\%        \\ \hline
\textbf{Ave. length}   & 212.50              & 244.95             & 19.55                & 18.23         \\ \hline
\textbf{Density}       & 0.2775\%            & 0.3003\%           & 0.0032\%             & 0.0005\%      \\ \hline
\end{tabular}
\end{table}

We evaluate the recommendation performance on an industrial Micro Video dataset and a public e-commerce dataset. The statistics of the datasets for our experiments are shown in Table~\ref{tbl:data_mv}. Appendix~\ref{appendix::dataset} illustrates the details of these two datasets.

\subsubsection{\textbf{Baselines and Evaluation Metrics}}
\begin{table*}[t!]

\caption{Performance comparisons for MAN on Micro Video dataset and Amazon dataset. } \label{tbl:realRes}
\begin{tabular}{cc|ccccccc|ccc|c}
\hline
\multirow{2}{*}{\textbf{Domain}} & \multirow{2}{*}{\textbf{Metric}} & \multicolumn{7}{c|}{\textbf{Single-domain}} & \multicolumn{4}{c}{\textbf{Cross-domain}} \\ \cline{3-13} 
 &  & DIN & Caser & GRU4REC & DIEN & SASRec & SLi-Rec & SURGE & NATR & PiNet & DASL & Ours \\ \hline
\multirow{4}{*}{\textbf{\begin{tabular}[c]{@{}c@{}}Micro \\ Video\\   A\end{tabular}}} & \textbf{AUC} & 0.5673 & 0.7744 & 0.7838 & 0.6666 & 0.7730 & 0.7558 & 0.7959 & {\ul 0.7972} & 0.7834 & 0.7879 & \textbf{0.8285} \\   
 & \textbf{MRR} & 0.5544 & 0.5740 & 0.5417 & 0.5264 & 0.5359 & 0.5337 & 0.5888 & {\ul 0.5899} & 0.5273 & 0.5568 & \textbf{0.6167} \\   
 & \textbf{NDCG} & 0.6628 & 0.6780 & 0.6531 & 0.6409 & 0.6488 & 0.6461 & 0.6892 & {\ul 0.6921} & 0.6422 & 0.6651 & \textbf{0.7112} \\   
 & \textbf{WAUC} & 0.7837 & 0.8053 & 0.7910 & 0.7654 & 0.7911 & 0.7729 & 0.8170 & {\ul 0.8197} & 0.7880 & 0.8075 & \textbf{0.8435} \\ \hline
\multirow{4}{*}{\textbf{\begin{tabular}[c]{@{}c@{}}Micro\\      Video\\      B\end{tabular}}} & \textbf{AUC} & 0.5613 & 0.7308 & 0.7625 & 0.6581 & {\ul 0.7794} & 0.7620 & 0.7605 & 0.7727 & 0.7595 & 0.7665 & \textbf{0.8094} \\   
 & \textbf{MRR} & 0.4526 & 0.4971 & 0.5285 & 0.4768 & {\ul 0.5472} & 0.5418 & 0.5042 & 0.5462 & 0.5037 & 0.5288 & \textbf{0.5756} \\   
 & \textbf{NDCG} & 0.5843 & 0.6184 & 0.6431 & 0.6025 & {\ul 0.6574} & 0.6529 & 0.6239 & 0.6571 & 0.6240 & 0.6431 & \textbf{0.6797} \\   
 & \textbf{WAUC} & 0.7246 & 0.7533 & 0.7860 & 0.7420 & {\ul 0.7957} & 0.7845 & 0.7645 & 0.7939 & 0.7705 & 0.7858 & \textbf{0.8215} \\ \hline
\multirow{2}{*}{\textbf{Domain}} & \multirow{2}{*}{\textbf{Metric}} & \multicolumn{7}{c|}{\textbf{Single-domain}} & \multicolumn{4}{c}{\textbf{Cross-domain}} \\ \cline{3-13} 
 &  & DIN & Caser & GRU4REC & DIEN & SASRec & SLi-Rec & SURGE & NATR & PiNet & DASL & Ours \\ \hline
\multirow{4}{*}{\textbf{\begin{tabular}[c]{@{}c@{}}Amazon\\      Video \\      Games\end{tabular}}} & \textbf{AUC} & 0.5577 & 0.5766 & 0.5303 & {\ul 0.6059} & 0.5234 & 0.5750 & 0.5975 & 0.5617 & 0.5740 & 0.5527 & \textbf{0.6559} \\   
 & \textbf{MRR} & 0.3736 & 0.3284 & 0.2953 & 0.3526 & 0.2833 & 0.3503 & {\ul 0.4667} & 0.3388 & 0.3419 & 0.3053 & \textbf{0.4755} \\   
 & \textbf{NDCG} & 0.5171 & 0.4854 & 0.4582 & 0.5046 & 0.4488 & 0.5009 & {\ul 0.5917} & 0.4918 & 0.4957 & 0.4667 & \textbf{0.5986} \\   
 & \textbf{WAUC} & 0.5587 & 0.5805 & 0.5395 & 0.6115 & 0.5257 & 0.5721 & {\ul 0.6311} & 0.5629 & 0.5847 & 0.5652 & \textbf{0.6686} \\ \hline
\multirow{4}{*}{\textbf{\begin{tabular}[c]{@{}c@{}}Amazon \\      Toys\end{tabular}}} & \textbf{AUC} & 0.6372 & 0.5138 & {\ul 0.6576} & 0.6321 & 0.5707 & 0.6106 & 0.6455 & 0.6127 & 0.5402 & 0.6237 & \textbf{0.6712} \\   
 & \textbf{MRR} & 0.5946 & 0.3293 & {\ul 0.5949} & 0.5669 & 0.3110 & 0.5292 & 0.5566 & 0.5070 & 0.3140 & 0.3515 & \textbf{0.6385} \\   
 & \textbf{NDCG} & 0.6879 & 0.4836 & {\ul 0.6889} & 0.6676 & 0.4721 & 0.6389 & 0.6602 & 0.6211 & 0.4729 & 0.5045 & \textbf{0.7221} \\   
 & \textbf{WAUC} & 0.6398 & 0.5058 & {\ul 0.6540} & 0.6421 & 0.5784 & 0.6184 & 0.6504 & 0.6217 & 0.5419 & 0.6305 & \textbf{0.6788} \\ \hline
\end{tabular}
\end{table*}

To demonstrate the effectiveness of our model, we compare it with two categories of competitive baselines: single-domain models and cross-domain models. Specifically, single-domain models are {DIN}~\cite{DIN} {Caser} ~\cite{Caser}, {GRU4REC}~\cite{GRU4REC}, {DIEN}~\cite{DIEN}, {SASRec}~\cite{kang2018self}, {SLi-Rec}~\cite{SLIREC} and {SURGE}~\cite{SURGE}. These single-domain models are trained on each domain independently following existing work~\cite{DASL, PiNet}. 

Besides, cross-domain models are {NATR}~\cite{NATR}, {PiNet}~\cite{PiNet} and {DASL}~\cite{DASL}. PiNet and DASL are adapted to our settings without fully-overlapped users (with the item sequence of another domain as empty). Other cross-domain models like MiNet~\cite{MiNet} and CoNet~\cite{CoNet} are not included in experiments because they are non-sequential models and will be much poor than sequential models~\cite{PiNet}.

All models are evaluated on two popular accuracy metrics AUC and GAUC~\cite{gunawardana_evaluating_2015}, as well as two ranking metrics, MRR and NDCG~\cite{SURGE}. 

\subsubsection{\textbf{Hyper-parameter Settings}}
The initial learning rate for Adam~\cite{Adam} is $0.001$ with Xavier initialization~\cite{xavier} to initialize the parameters. Regularization coefficients are  searched in $[1 e^{-7}$, $1 e^{-5}$, $1 e^{-3}]$. The batch size is set as 200 and 20, respectively, for the Micro Video dataset and Amazon dataset. The embedding sizes of all models with 40 and 20 are fixed for the Micro Video dataset and Amazon dataset, respectively. MLPs with layer size $[100,64]$ and $[20,10]$ are exploited for the prediction layer on the Micro Video dataset and Amazon dataset, respectively. Item sequence length of 250 is set for the Micro Video dataset, and 20 is set for the Amazon dataset. The numbers of group prototypes are searched from [1, 5, 10, 20].

\subsection{Overall Performance (RQ1)}

The performance comparisons over all models are as shown in Table~\ref{tbl:realRes}, where SASRec and SURGE with better performance are leveraged as backbones on these two datasets, respectively. It can be observed that:
\begin{itemize}[leftmargin=*]
\item \textbf{Our approach performs best}. Our model MAN significantly outperforms all baselines under all metrics. Specifically, our model improves AUC against all baselines by 4.10\% and 3.85\% on Micro Video A and Micro Video B, respectively, while by 8.25\% and 2.07\% on Amazon Video Games and Amazon Toys, respectively. In general, the improvement is more consistent across evaluation metrics on the Micro Video dataset with more overlapped users. The Amazon dataset with extremely sparse data sees the highest improvement (8.25\%), which verifies that our approach can address the sparse data problem, promoting the sequential learning of both domains simultaneously and that of less interacted domains even more sharply. 


\item \textbf{Existing cross-domain sequential recommenders rely heavily on overlapped users or items}. PiNet and DASL are based on fully-overlapped user datasets~\cite{PiNet, DASL}, but they are indeed comparable with GRU4REC under datasets without fully-overlapped users, either outperforming or even underperforming. In contrast to them, our proposed approach outperforms all baselines and improves the backbones significantly, which illustrates the effectiveness of our cross-domain modeling without user overlapping. Though NATR achieves decent performance on the Micro Video dataset with a lot of overlapped items, it fails to achieve effective cross-domain modeling on the Amazon dataset with limited overlapped items.

\item \textbf{Sequential recommenders are effective but with data sparsity bottleneck}. 
Based on the Micro Video dataset, comparing the sequential models (i.e., Caser, GRU4REC, DIEN, SASrec, SLi-Rec, and SURGE) with the non-sequential model (i.e., DIN), it is necessary for us to model the chronological relationship between items. Besides, SASRec and SURGE are comparable and outperform all other single-domain sequential models, which illustrates the capacity of self-attention to handle long-term information and verifies the effectiveness of compressing information with metric learning. The observation of sequential models is consistent with the experimental results of the SURGE~\cite{SURGE} paper. 
Based on the Amazon dataset, DIN even outperforms some sequential models, i.e., SASRec, which also drops a lot under such a short sequence scene. Though sequential models are the potential for capturing the chronological relationship between items, they are blocked by the data sparsity. 
Besides we have also attempted to train them with two domains simultaneously ("Shared" models in the backbone study), but the results show that one domain's optimization will have a negative impact on another domain, leading to optimization conflict. Thus it is necessary to design cross-domain modeling to avoid optimization conflict and negative transfer.
\end{itemize}

\subsection{Impact of Each Component (RQ2)}

\begin{table}[t!]
\caption{Ablation study of the proposed components on Micro Video dataset and Amazon dataset.}\label{tbl:ablation_component}

\begin{tabular}{cccccc}
\hline
\textbf{Domain} & \textbf{Model} & \textbf{AUC} & \textbf{MRR} & \textbf{NDCG@10} & \textbf{WAUC} \\ \hline
\multirow{4}{*}{\begin{tabular}[c]{@{}c@{}}Micro\\ Video\\ A\end{tabular}} & w/o ISA & {\ul 0.8136} & 0.5795 & 0.6826 & 0.8216 \\ \cline{2-6} 
 & w/o SFA & 0.8133 & {\ul 0.5895} & {\ul 0.6904} & {\ul 0.8292} \\ \cline{2-6} 
 & w/o GPA & 0.7983 & 0.5714 & 0.6762 & 0.8134 \\ \cline{2-6} 
 & w all & \textbf{0.8285} & \textbf{0.6167} & \textbf{0.7112} & \textbf{0.8435} \\ \hline
\multirow{4}{*}{\begin{tabular}[c]{@{}c@{}}Micro \\ Video\\ B\end{tabular}} & w/o ISA & {\ul 0.8059} & {\ul 0.5644} & {\ul 0.6711} & {\ul 0.8162} \\ \cline{2-6} 
 & w/o SFA & 0.7939 & 0.5559 & 0.6643 & 0.8073 \\ \cline{2-6} 
 & w/o GPA & 0.7996 & 0.5631 & 0.6701 & 0.8147 \\ \cline{2-6} 
 & w all & \textbf{0.8094} & \textbf{0.5756} & \textbf{0.6797} & \textbf{0.8215} \\ \hline
\multirow{4}{*}{\begin{tabular}[c]{@{}c@{}}Amazon\\ Video\\ Games\end{tabular}} & w/o ISA & 0.6195 & {\ul 0.4437} & {\ul 0.5743} & 0.6352 \\ \cline{2-6} 
 & w/o SFA & {\ul 0.642} & 0.4426 & 0.5735 & {\ul 0.6523} \\ \cline{2-6} 
 & w/o GPA & 0.6195 & 0.4198 & 0.5549 & 0.6288 \\ \cline{2-6} 
 & w all & \textbf{0.6559} & \textbf{0.4755} & \textbf{0.5986} & \textbf{0.6686} \\ \hline
\multirow{4}{*}{\begin{tabular}[c]{@{}c@{}}Amazon \\ Toys\end{tabular}} & w/o ISA & 0.6499 & 0.5497 & 0.6548 & 0.6516 \\ \cline{2-6} 
 & w/o SFA & 0.6502 & 0.6126 & 0.7021 & 0.6583 \\ \cline{2-6} 
 & w/o GPA & {\ul 0.6546} & {\ul 0.6183} & {\ul 0.7074} & {\ul 0.6606} \\ \cline{2-6} 
 & w all & \textbf{0.6712} & \textbf{0.6385} & \textbf{0.7221} & \textbf{0.6788} \\ \hline
\end{tabular}
\end{table}

 \begin{figure*}[t!]
	\begin{center}
	\setlength\tabcolsep{2pt}   

		\begin{tabular}{cccc}
			\includegraphics[width=.5\columnwidth]{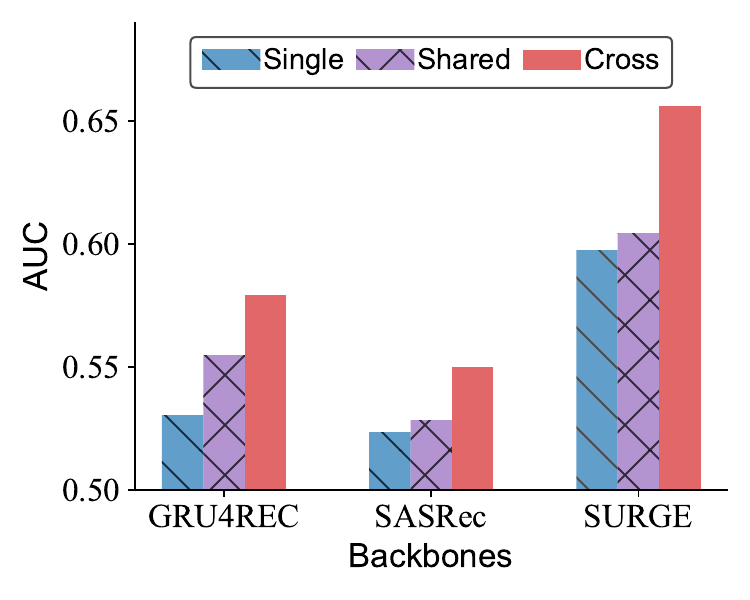} & 	\includegraphics[width=.5\columnwidth]{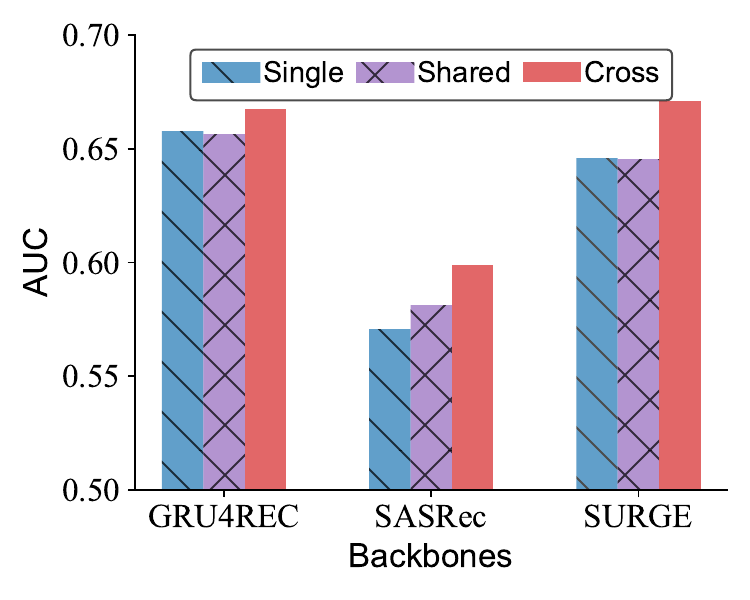} & \includegraphics[width=.5\columnwidth]{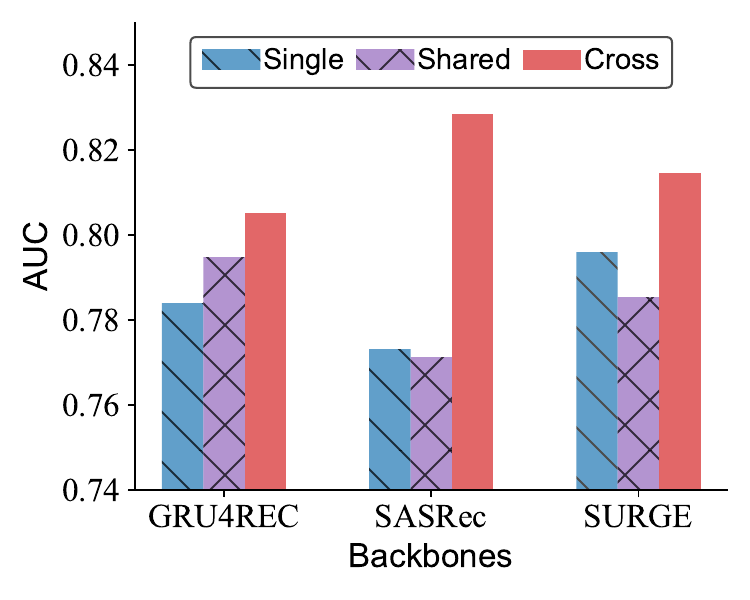} & \includegraphics[width=.5\columnwidth]{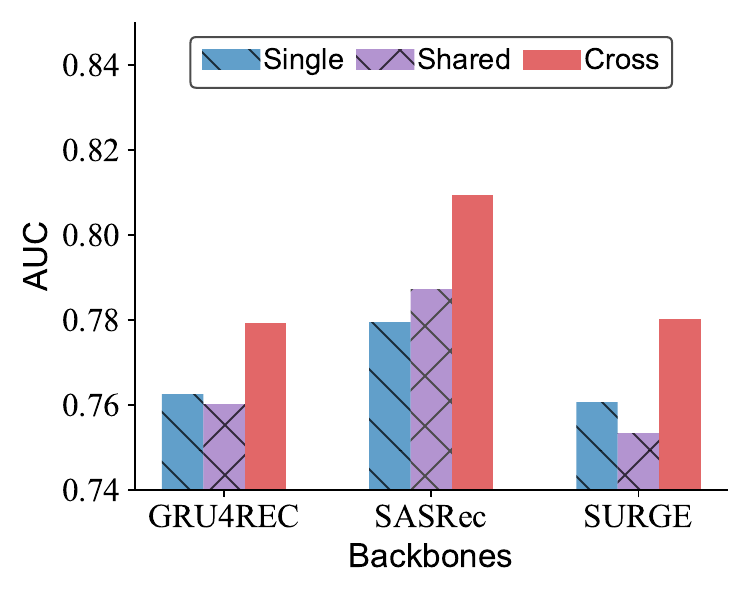} \\
		(a) Micro Video A & (b) Micro Video B & (c) Video Games & (d) Toys
			
		\end{tabular}
	\end{center}
	\caption{{AUC performance of MAN with different backbones on Micro Video dataset and Amazon dataset. Here "Single" means backbone models trained with single domain data, which refers to the local module. "Shared" means shared backbone model trained with cross-domain data, which refers to the global module. "Cross" is the backbone equipped with our method.}}	\label{fig:backbone}
\end{figure*} 

\begin{figure*}[t!]
	\begin{center}
		\begin{tabular}{cccc}
			\includegraphics[width=.45\columnwidth]{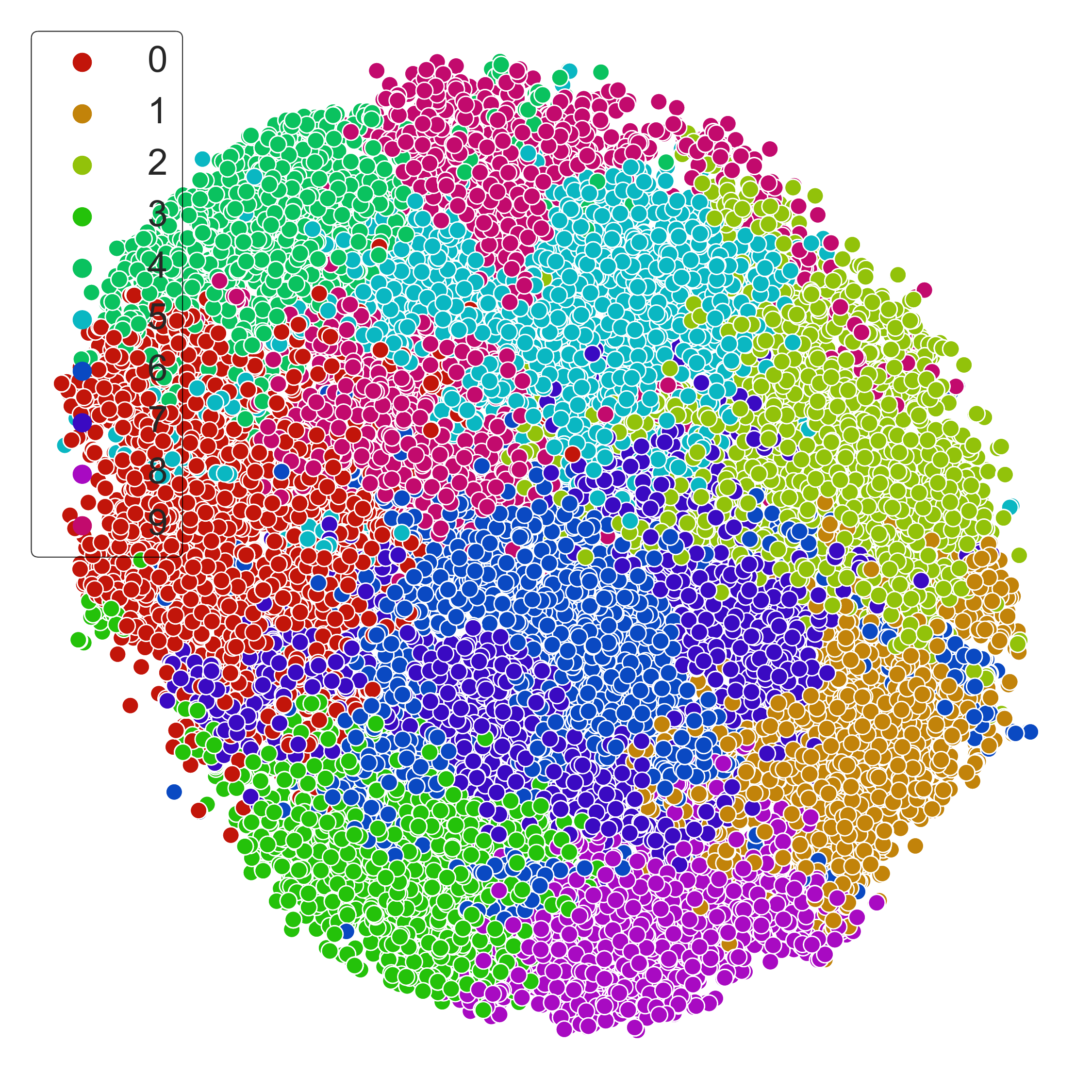}  &
		\includegraphics[width=.45\columnwidth]{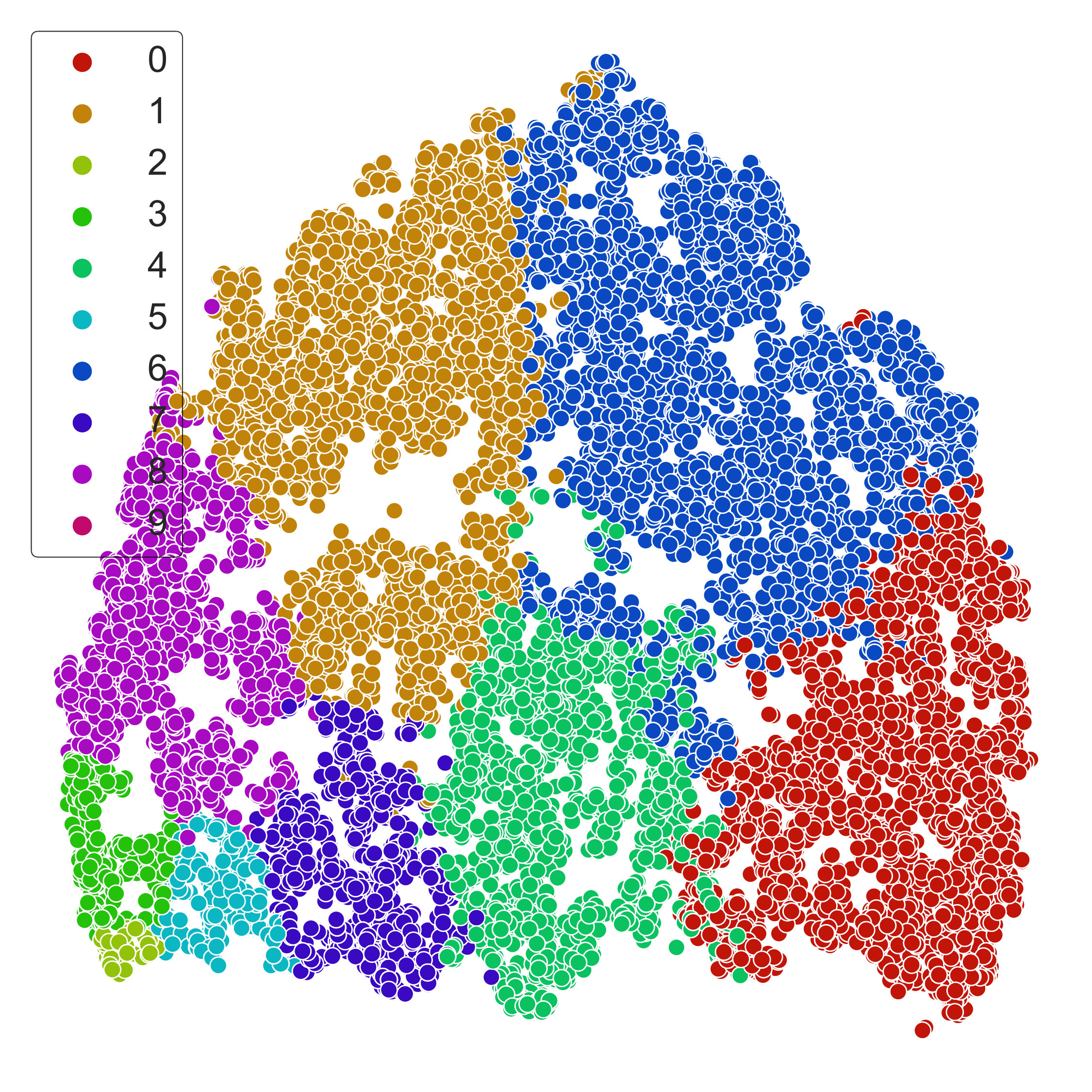}
		&
		\includegraphics[width=.45\columnwidth]{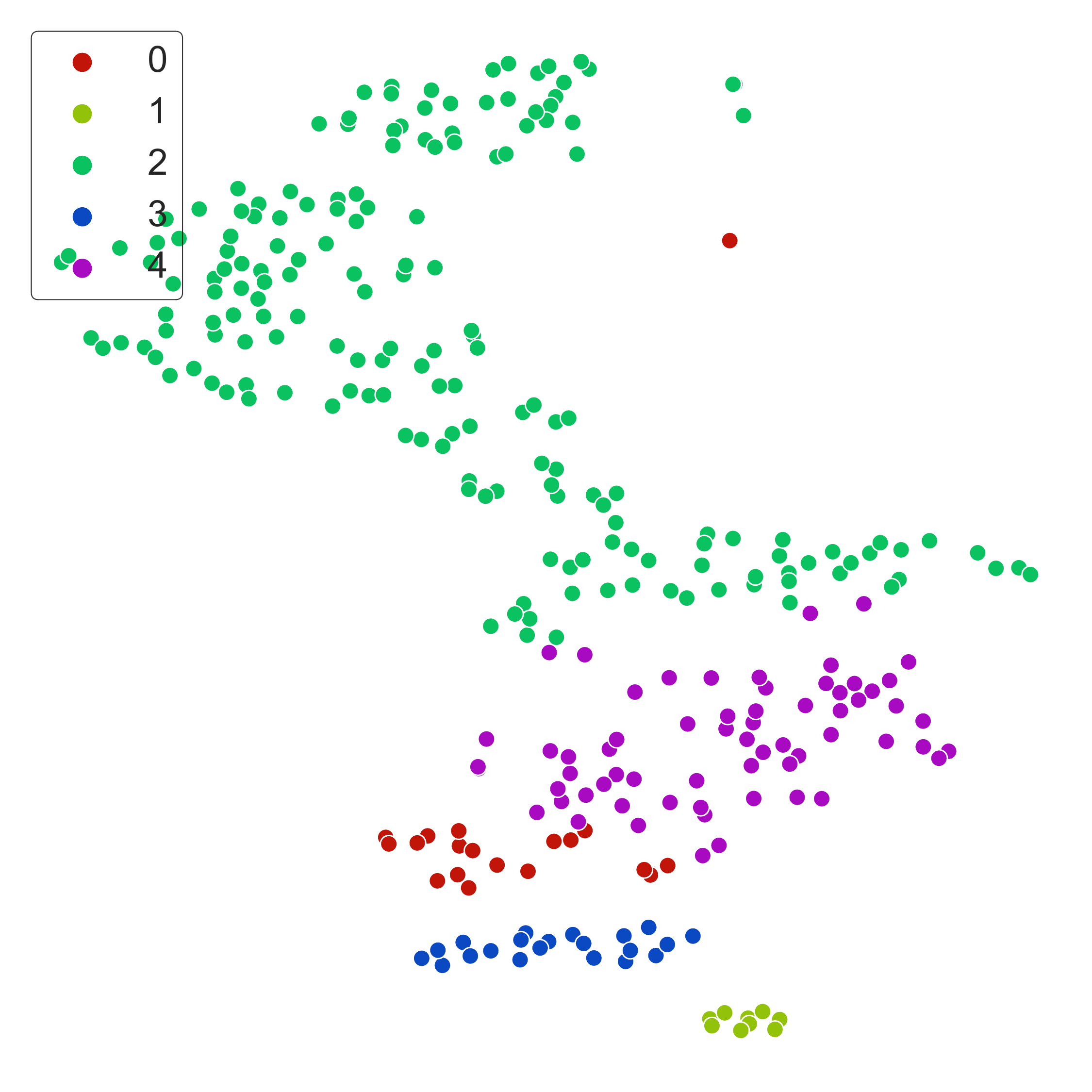}
		&
		\includegraphics[width=.45\columnwidth]{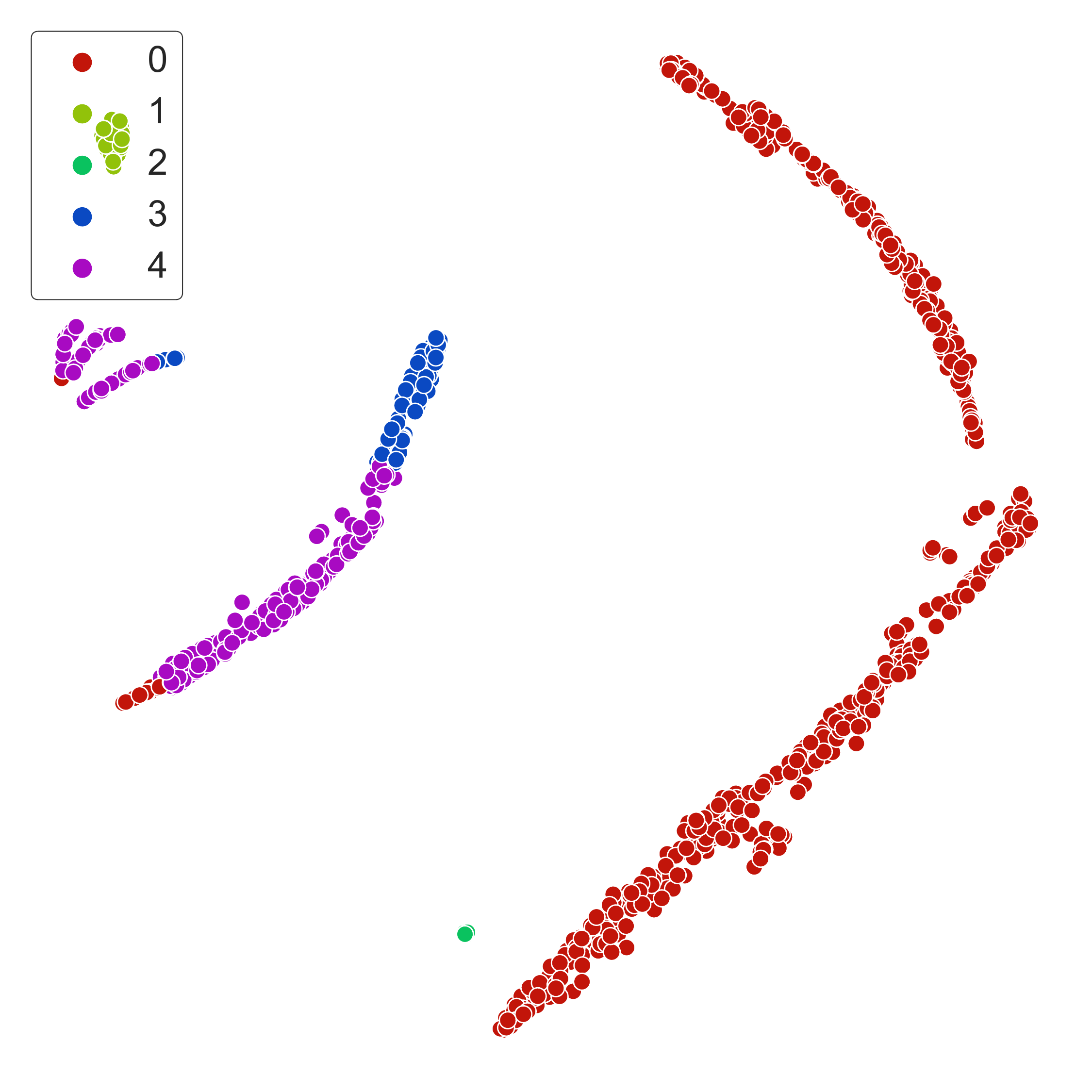} \\
		(a) Micro Video A & (b) Micro Video B & (c) Video Games & (d) Toys
		\end{tabular}
	\end{center}
	\caption{K-Means and t-SNE visualization of pooled group representations on Micro Video dataset and Amazon dataset, with different colors representing different groups. Group patterns across domains of two datasets are captured by the different distribution of group representations. (Best view in color.)}	\label{fig:group_weight}
\end{figure*}
To study the impact of our proposed components, we compare our model with that detaching Item Similarity Attention (ISA) module, Sequence-fusion Attention (SFA) module, and Group-prototype Attention (GPA) module on two datasets under four evaluation metrics, as shown in Table~\ref{tbl:ablation_component}. Firstly, it can be observed that the shared group prototypes of {GPA} are {most effective} in both Micro Video and Amazon Video datasets, illustrating that there are similar interest groups across different domains. Besides, the performance also drops a bit when removing the sequence-fusion component (most effective in Micro Video B), i.e., SFA for fusing the local and global sequential patterns, which means there are truly common sequential patterns across different domains. There are also similar items across different domains when the performance decreases after the detaching item similarity attention module (most effective in Amazon Toys). 

In short, Group-prototype Attention is the most important among the three proposed attentions.

\subsection{Backbone Study (RQ3)}

{Here, we study whether GRU4REC, SASRec, and SURGE can be boosted under our proposed method. That is to say, whether our proposed method is model-agnostic. The reason why we choose these three models is that they perform better on the experimented datasets. Figure~\ref{fig:backbone} shows the results of our method with different backbones on two datasets under AUC evaluation, where we can observe that:
\begin{itemize}[leftmargin=*]
\item \textbf{Our proposed method is model-agnostic}. The selected backbones are all boosted by our proposed MAN, which means our proposed method is model-agnostic. The backbones selected here are RNN-based, attention-based, and even graph-based models. Thus our method can be applied in various state-of-the-art sequential recommendation models to boost their performance. 
\item \textbf{Our proposed method performs better on larger datasets}. 
The improvement on the Micro Video dataset is generally more obvious than that on the Amazon dataset. This is because a large dataset can provide rich cross-domain information. 
\end{itemize}}

\subsection{User Group Visualization (RQ4)}

In this section, the embeddings of all users' pooled group representations will be visualized to show the patterns our group-prototype attention module has captured. 

The pooled group representation (calculated in Eqn.\eqref{eq:group_repre_a}) for each user is visualized with K-Means and t-SNE, as shown in Figure~\ref{fig:group_weight}. More specifically, we first apply K-Means on the pooled group representations to cluster data into $N_g$ groups. Then t-SNE is exploited to reduce the group representations into two-dimensional space, and the clustered groups by K-Means are used to label each user. It can be observed that: 
(1) for each dataset, the group patterns vary across different domains, where the users under Micro Video A are distributed evenly while the users under Micro Video B mostly belong to groups 0, 1, and 6. On the Amazon dataset, the users mostly belong to group 2, and group 0 under Video Games and Toys, respectively; (2) for two datasets, the users on the Amazon dataset are distributed more unbalanced and dispersed than those on the Micro Video dataset, which may because the Amazon dataset is more sparse.

\vspace{0.1cm}
\section{Related Work}
\vspace{0.1cm}

There are two fields of work related to our proposed model: sequential recommendation and cross-domain recommendation..

\vspace{0.2cm}

\noindent \textbf{Sequential Recommendation}
Sequential Recommendation~\cite{SRs} is the fundamental model of our work, which models the user's historical behaviors as a sequence of time-aware items, aiming to predict the probability of the next item.
Initially, the Markov chain is exploited to model the sequential pattern of item sequence as FPMC~\cite{rendle2010factorizing}. 
To further extract the high-order interaction between the historical items, researchers have also applied deep learning models such as recurrent neural network~\cite{GRU, LSTM}, convolution neural network~\cite{CNN} and attention network~\cite{vaswani2017attention} in recommender systems~\cite{GRU4REC, DIEN, Caser, kang2018self, DIN}.
However, recurrent neural network-based and convolution neural network-based methods often pay attention to the recent items before the next item, failing to model the long-term interest.
Recently, researchers have also combined the sequential recommendation model and traditional recommendation model such as matrix factorization~\cite{koren2009matrix} to model the long and short-term interest~\cite{SLIREC, zhao2018plastic} while SURGE~\cite{SURGE} exploits metric learning to compress the item sequence. Some recent works like DFAR~\cite{lin2023dual} and DCN~\cite{lin2022dual} focus on capturing more complex relations behind sequential recommendation.
In this paper, we perform cross-domain learning based on sequential recommendation models to achieve knowledge transfer between different domains.
\vspace{0.2cm}

\noindent \textbf{Cross-Domain Recommendation} Cross-domain recommender systems~\cite{CDR} are an effective solution to the highly sparse data problem and cold-start problem that sequential recommendation meets. 
Early cross-domain recommendation models are based on single-domain recommendation, assuming that auxiliary user behaviors across different domains will benefit the target domain's user modeling~\cite{singh_relational_2008, hu_personalized_2013, loni2014cross}.
Indeed, the most popular approaches are often based on transfer learning~\cite{Transfer} to transfer the user embedding or item embedding from the source domain to improve the target domain's modeling, including MiNet~\cite{MiNet}, CoNet~\cite{CoNet} and itemCST~\cite{itemCST} etc. 

However, industrial platforms tend to improve all domains of their products simultaneously instead of improving the target domain without consideration of the source domain. Thus, dual learning~\cite{long_dual_2012, he_dual_2016}, which can achieve simultaneous improvements across both source domain and target domain, grabs researchers' attention and has already been applied in cross-domain recommender systems~\cite{DTCDR, DDTCDR}. 
Moreover, to enhance the recommendation performance across all domains simultaneously, researchers have proposed some dual-target approaches focusing on sequential modeling~\cite{PiNet, DASL, chen2021dual}, which addresses the sparse data problem and cold-start problem promisingly and considers the performance of both source domain and target domain.
Specifically, PiNet~\cite{PiNet} tackles the shared account problem and transfers account information from one domain to another domain where the account also has historical behaviors; DASL~\cite{DASL} proposes dual embedding to interact embeddings and dual attention to mix the sequential patterns for the same users across two domains.
Besides PiNet and DASL, DAT-MDI~\cite{chen2021dual} applies dual attention like DASL on session-based recommendation without relying on user overlapping. However, requiring the item sequence pairs in two domains as input is unreasonable because the item sequences of two domains are often independent of each other despite belonging to the same user.
Hence such a dual attention manner by mixing the sequence embedding of two domains will not result in a promising performance, theoretically speaking, under a non-overlapped user scene. Though NATR~\cite{NATR} tends to avoid user overlapping, it is a non-sequential and single-target model.

In this paper, we perform cross-domain learning in a dual-target manner to achieve simultaneous improvements across different domains without any prior assumption of overlapped users or items.

\vspace{0.2cm}

\section{Conclusions and Future Work}

In this work, we studied the task of sequential recommender systems in a cross-domain manner from a more practical perspective without any prior assumption of overlapped users. Such exploration brought us three key challenges from the item, sequence, and group levels. 
To address these three challenges, we proposed a novel solution named MAN with local and global modules, mixing three attention networks and transferring at the group level. The first one was the local/global encoding layer that captures the sequential pattern from domain-specific and cross-domain perspectives. Secondly, we further proposed the item similarity attention that captured the similarity between local/global item embeddings and target item embedding, the sequence-fusion attention that fused sequential patterns across global encoder and local encoder, and the group-prototype attention with several group prototypes to share the sequential user behaviors implicitly without leveraging the user ID. Finally, we proposed a local/global prediction layer to evolve the domain-specific and cross-domain interests.

As for future work, we plan to conduct online A/B tests to further evaluate our proposed solution's recommendation performance in the real-world product. 
We also consider applying MAN with more advanced sequential backbones, even from other fields, to explore the generalization of our proposed modules.

\clearpage
\bibliographystyle{ACM-Reference-Format}
\bibliography{sample-base}
\clearpage
\appendix
\section{APPENDIX FOR REPRODUCIBILITY}
\subsection{Notation}
We present all used symbols as Table~\ref{tbl:notation} for clearer understanding.

\begin{table}[hbt!]
	\caption{Notation table of important symbols.}
	\label{tbl:notation}
\small
		\begin{centering}
			\setlength\tabcolsep{4pt}   

			\begin{tabular}{ l|l} \toprule
			\textbf{Notations} & \textbf{Descriptions} \\ \hline
				$N_g$ &  number of groups\\
$T$ & maximum length of item sequence \\

				$\mathcal{I}^{A}, \mathcal{I}^{B}$ & item sets in  domain A and B \\
				$i^{A}_{t} \in \mathcal{I}^{A}$, $i^{B}_{t} \in \mathcal{I}^{B}$ & the $t$-th item clicked by given users in domain A and B  \\
				\hline
				$\mathcal{R}^A$, $\mathcal{R}^B$ & training sets of domain A and domain B\\
								
				$y^A_{u, t} \in \{0,1\}$ & 
				$			\begin{cases}
					1&  \text{if user in domain A}  \text{ clicks item } i^A_t \\
					0&  \text{if user in domain A}  \text{ does not click item } i^A_t 
				\end{cases}$ \\	
								$y^B_{u, t} \in \{0,1\}$ & 
				$			\begin{cases}
					1&  \text{if user in domain B} \text{ clicks item } i^B_t \\
					0&  \text{if user in domain B} \text{ does not click item } i^B_t 
				\end{cases}$ \\

				\hline
				$D, D'$ & number of latent dimensionality \\

	$\mathbf{M}^A \in \mathbb{R}^{|\mathcal{I}^A | \times D}$ & item embedding matrix for domain A\\
	$\mathbf{M}^B \in \mathbb{R}^{|\mathcal{I}^B | \times D}$ &  item embedding matrix for domain B \\

				$\mathbf{M} \in \mathbb{R}^{|\mathcal{I}^A \cup \mathcal{I}^B| \times D'}$ & item embedding matrix for both domains\\
				 $\mathbf{P}^A$, $\mathbf{P}^B \in \mathbb{R}^{T \times D}$, &  position embedding matrix for A and B \\
				 $\mathbf{P} \in \mathbb{R}^{T \times D'}$ &  positional embedding matrices for both domains\\
			
				$\mathbf{G} \in \mathbb{R}^{N_g \times D}$ & group prototype embedding matrix\\

				\bottomrule
			\end{tabular}
			
		\end{centering}
	
\end{table}

\subsection{\textbf{Group Number Study} (RQ5)}\label{appendix::group_num}
\begin{figure}[hbt!]
	\begin{center}
	\setlength\tabcolsep{2pt}   

		\begin{tabular}{cc}
			\includegraphics[width=.48\columnwidth]{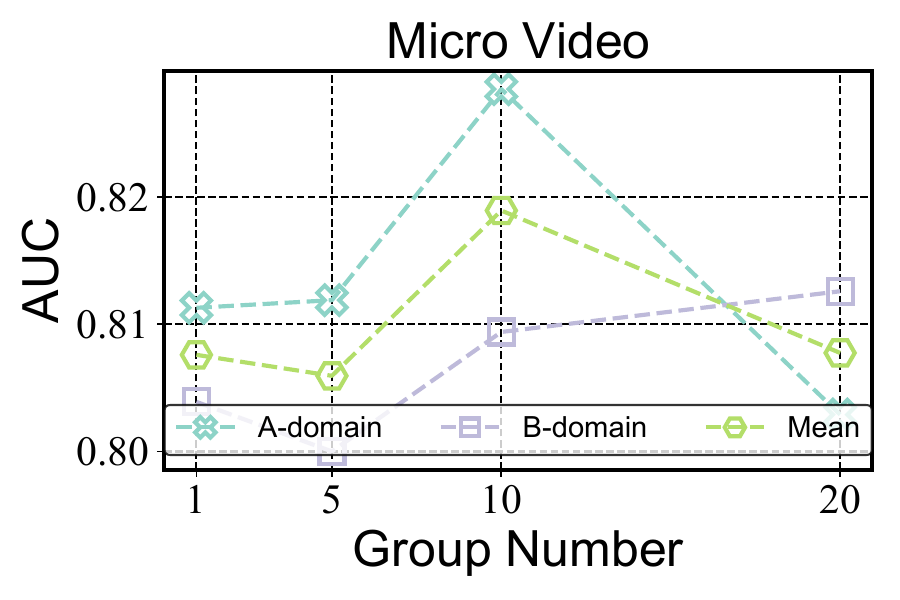} & \includegraphics[width=.48\columnwidth]{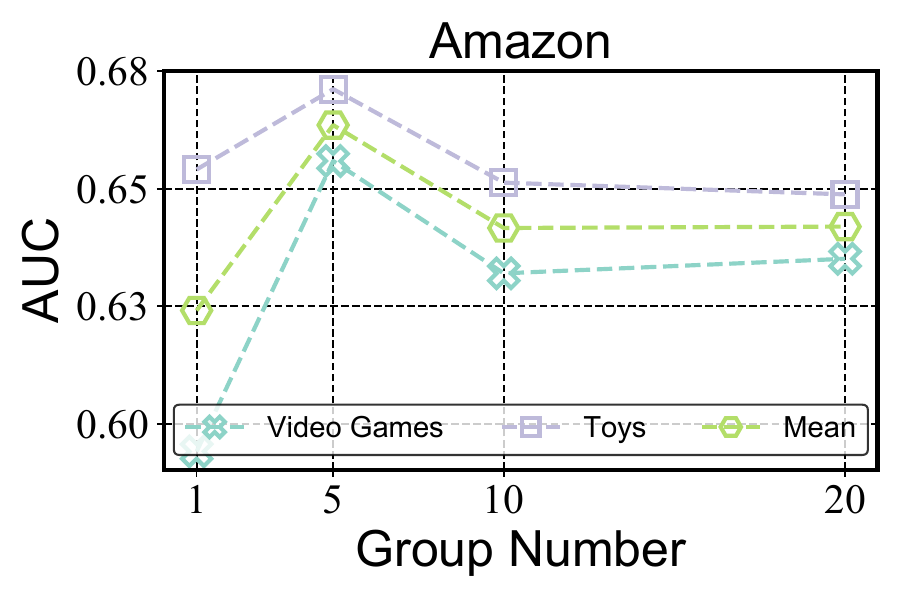} 
		\end{tabular}
	\end{center}
	\caption{Recommendation performance of MAN \textit{w.r.t.} the number of groups on Micro Video dataset and Amazon dataset. Here Mean is the average AUC performance of two domains. Group number varies from 1 to 20.}	\label{fig:group_num}
	\vspace{-0.4cm}
\end{figure}

We vary the number of groups from $\{1, 5, 10 ,20\}$ as Figure~\ref{fig:group_num} where AUC is tested to explore the best number of groups. From Figure~\ref{fig:group_num}, we can observe that: (1) for the Micro Video dataset, AUC reaches the peak when the number of groups is 10 under A domain, and AUC is best at group number 20 under B domain, while the mean value of AUC is best at 10 for these two domains; (2) for Amazon dataset, AUC is best at group number 5 for both domains.

\subsection{Implementation}
All the models are implemented based on Python with a TensorFlow~\footnote{ https://www.tensorflow.org} framework of Microsoft~\footnote{https://github.com/microsoft/recommenders}. Besides, we also exploited Python to perform K-Means and t-SNE on group representation for each user. The codes for our model and visualization are available on Github~\footnote{https://github.com/KDD-334/MAN} with processed Amazon dataset. The K-Means and t-SNE visualization code and embedding files to be visualized is under the directory ``\url{MAN/Code-visualization}''. Note that we will release the Micro Video dataset to benefit the community in the future.

Each item embedding is concatenated with a domain embedding according to the specific domain of input items. To avoid the distortion on the local and global sequential learning, we also stop the back propagation of $\mathbf{S}^A$, $\mathbf{S}^B$ and $\mathbf{S}^{A_g}$, $\mathbf{S}^{B_g}$ in Sequence-fusion Attention module of Section~\ref{sec:chronological}, which has been verified to be more effective by our early attempt. Besides, the back propagations of $\mathbf{S}^A$ and $\mathbf{S}^B$ are also stopped in Group-prototype Attention module of Section~\ref{sec:tsa}.

MLP for SASRec backbone is a MLP layer sandwiched two normalization layers. For the SURGE~\cite{SURGE} backbone, we use the same input as the paper for the local prediction layer and global prediction layer, respectively. Besides, we concatenate the outputs of our item similarity attention module, sequence-fusion attention module and group-prototype attention module to the input of local prediction layer.

\noindent \textbf{Single-domain Models}

\begin{itemize}[leftmargin=*]

    \item \textbf{DIN}~\cite{DIN}: It represents the user by the aggregation of the historical items based on the attention weights calculated via querying the target item with the historical items. 
    \item \textbf{Caser} ~\cite{Caser}: It performs convolution filters on the historical item embedding to capture the sequential pattern.
    \item \textbf{GRU4REC}~\cite{GRU4REC}: It models session sequence and represents user preference by the final state based on GRU~\cite{GRU}.
    
    \item \textbf{DIEN}~\cite{DIEN}: It proposes an interest extraction GRU layer and interest evolution GRU layer to capture the sequential pattern.
    
\item \textbf{SASRec}~\cite{kang2018self}: It captures the sequential pattern via hierarchical self-attention network.
\item \textbf{SLi-Rec}~\cite{SLIREC}: It proposes an attention framework and improves LSTM with time awareness to jointly model the long and short-term interests.

\item \textbf{SURGE}~\cite{SURGE}: It first exploits metric learning to construct a tight item-item interest graph for historical item sequence before performing cluster-aware and query-aware graph convolutional propagation and graph pooling.
\end{itemize}
\textbf{Cross-domain Models}
\begin{itemize}[leftmargin=*]
\item \textbf{NATR}~\cite{NATR}: It relies on the overlapped items and performs linear transformation to transfer the item representation from the source domain to improve the performance in the target domain. 

\item \textbf{PiNet}~\cite{PiNet}: It represents the user by a shared account filter unit, transfers user information via a cross-domain transfer unit, and encodes the sequence by GRU.
\item \textbf{DASL}~\cite{DASL}: It is the state-of-art cross-domain sequential model proposing dual embedding to represent the cross-domain user and dual attention to model the cross-domain sequential pattern.
\end{itemize}

\subsection{Datasets and Evaluation Metrics}~\label{appendix::dataset}
The public Amazon dataset is available here~\footnote{\url{http://jmcauley.ucsd.edu/data/amazon/index_2014.html}} and we also have uploaded the filtered dataset after 10-core setting on the Github of the code and the supplementary material. The statistics of our adopted Micro Video dataset and Amazon dataset before filtering by 10-core settting are as Table~\ref{tbl:data_before}. The detailed illustration of them are as below.
\begin{itemize}[leftmargin=*]
\item \textbf{Micro Video}. This dataset contains two domains, Micro Video A and B, collected from one of the largest micro-video platforms in China, where users can share their videos. User behaviors such as click, like, follow (subscribe), and forward are recorded in the dataset. We downsample the logs from September 11 to September 22, 2021, and filter out inactive users and videos via the 10-core setting~\cite{SURGE}. We split the behaviors before 12 pm on the last day and after 12 pm on the last day, respectively, as the validation set and test set. Other behaviors are used for training.

\item \textbf{Amazon}~\footnote{Amazon.com}. This highly sparse dataset with two domains is adopted by the existing cross-domain sequential recommendation work DASL~\cite{DASL}, with few overlapped items but some overlapped users. We treat all the rating records as implicit feedback, also with the 10-core setting. The datasets include records from May 1996 to July 2014. We split the behaviors before June of the last year and after June of the last year, respectively, as the validation set and test set. Other behaviors are used for training.
\end{itemize}

The description of our adopted metrics is listed as:
\begin{itemize}[leftmargin=*]
     \item \textbf{AUC} calculates the probability that the predicted positive target item's score is ranking higher than the predicted negative item's score, evaluating the model's accuracy of classification performance.
     \item \textbf{GAUC} is a weighted average of each user's AUC, where the weight is his/her click number. It evaluates the model performance in a more bias-aware and fine-grained manner.
     \item \textbf{MRR} is the mean reciprocal rank, which averages the value of the first hit item's inverse ranking.
     \item \textbf{NDCG@K} thinks highly of those items at higher positions in the recommended K items, where the test items rank higher will result in better evaluating performance. In our experiments, $K$ is set to 10, a popular setting in related work~\cite{kang2018self}.
 \end{itemize}
 
\subsection{Parameter Settings}

\begin{table}[t!]
\caption{Data statistics for Micro Video dataset and Amazon dataset before being filtered by 10-core setting.}\label{tbl:data_before}
\setlength\tabcolsep{4pt}   
\begin{tabular}{c|c|c|c|c}
\hline
Dataset   & \multicolumn{2}{c|}{Micro Video}                             & \multicolumn{2}{c}{Amazon}                                 \\ \hline
Domain    & \multicolumn{1}{c|}{A-domain} & \multicolumn{1}{c|}{B-domain} & \multicolumn{1}{c|}{Video Games} & \multicolumn{1}{c}{Toys} \\ \hline
\#Users   & 43,919                       & 37,692                       & 826,767                         & 1,342,911                \\ \hline
\#Items   & 147,813                      & 131,732                      & 50,210                          & 327,698                  \\ \hline
\#Records & 18,011,737                   & 14,908,625                   & 1,324,753                       & 2,252,771                \\ \hline
\end{tabular}
\end{table}

All models are trained with 2 steps for early stop.

Activated by RELU (Rectified Linear Unit), MLP with layer size $[80,40]$ and $[32,16]$ are exploited for the Item Similarity Attention module on Micro Video dataset and Amazon dataset, respectively. 
 
 For Micro Video dataset and Amazon dataset, the dimensions of domain embeddings are set as 8 and 4 while those of item embeddings are set as 32 and 16, respectively.
 
 For SURGE backbone, we set the parameters following the paper.
 For the comparison methods PiNet~\footnote{https://github.com/mamuyang/PINet} and DASL~\footnote{https://github.com/lpworld/DASL}, we implement it under our framework based on the source code provided by the authors and can be referred in the footnotes. For DASL baseline, we do not pre-train the model as the paper for fair comparison and when we directly execute their provided code, we get poorer performance than the results in their paper under Amazon dataset.

 Other parameters of our model with SURGE backbone can refer to ``gcn.yaml'' under path ``\url{MAN/reco_utils/recommender/deeprec/config/}''.

\end{document}